\documentclass[aps,floatfix,nofootinbib,showpacs,twocolumn,superscriptaddress]{revtex4}

\usepackage{graphicx}
\usepackage{dcolumn}
\usepackage{amsmath}
\usepackage{longtable}

\newcommand{\sfrac}[2]{\mbox{\footnotesize $\frac{#1}{#2}$}} 

\begin{document}

%\title{Bethe-Salpeter equation and the dressed quark-gluon vertex}
\title{Sketching the Bethe-Salpeter kernel}

\author{Lei Chang}
\affiliation{Institute of Applied Physics and Computational Mathematics, Beijing 100094, China}

\author{Craig D. Roberts}
%\email[Corresponding author: ]{cdroberts@anl.gov}
\affiliation{Physics Division, Argonne National Laboratory, Argonne,
Illinois 60439, USA}
\affiliation{Department of Physics, Peking University, Beijing 100871, China}
\affiliation{School of Physics, The University of New South Wales, Sydney NSW 2052, Australia}

%\date{\today}

\begin{abstract}
An exact form is presented for the axial-vector Bethe-Salpeter equation, which is valid when the quark-gluon vertex is fully dressed.  A Ward-Takahashi identity for the Bethe-Salpeter kernel is derived therefrom and solved for a class of dressed quark-gluon vertex models.  The solution provides a symmetry-preserving closed system of gap and vertex equations. The analysis can be extended to the vector equation.  This enables a comparison between the responses of pseudoscalar- and scalar meson masses to nonperturbatively dressing the quark-gluon vertex.  The result indicates that dynamical chiral symmetry breaking enhances spin-orbit splitting in the meson spectrum.
\end{abstract}


\maketitle

%\section{Introduction}
Understanding the spectrum of hadrons with masses less than 2\,GeV is an essential step toward revealing the essence of light-quark confinement and dynamical chiral symmetry breaking (DCSB) and describing hadron structure in terms of QCD's elementary degrees of freedom.  These are basic questions, which define a frontier of contemporary hadron physics, yet there are no reliable Poincar\'e invariant calculations of this spectrum.

In this spectrum the $\pi(1300)$ is a radial excitation of the $\pi(140)$ \cite{Holl:2004fr,Holl:2005vu}, the $\pi(1800)$ is possibly a hybrid \cite{Barnes:1996ff}, and the dressed-quarks within scalar- and pseudovector-mesons possess significant orbital angular momentum \cite{Burden:1996nh,Ackleh:1996yt,Bloch:1999vka}.  Hence, relative to ground-state pseudoscalar and vector mesons, these states are sensitive to different features of the interaction between light-quarks and to that interaction at larger distances \cite{Ackleh:1996yt,Bloch:1999vka,Holl:2004fr,Holl:2005vu}.  Such systems are therefore more responsive to the dynamics of light-quark confinement.   The large magnitude of both the $\pi$-$\rho$ mass difference and the splitting between parity partners are two consequences of DCSB, which materially influences the hadron spectrum.  It is anticipated but not proven that confinement is sufficient to ensure DCSB.  However, the reverse is not true \cite{Roberts:2007jh,Roberts:2007ji}. 

In connection with confinement it is important to appreciate that the static potential measured in numerical simulations of quenched lattice-regularised QCD is not related in any simple way to the question of light-quark confinement.  It is a basic feature of QCD that light-quark creation and annihilation effects are essentially nonperturbative and therefore it is impossible in principle to compute a potential between two light quarks \cite{Bali:2005fu}.  

Confinement can be related to the analytic properties of QCD's Schwinger functions \cite{Roberts:2007ji,Krein:1990sf,Roberts:1994dr}.  
From this standpoint the question of light-quark confinement can be translated into the challenge of charting the infrared behavior of QCD's \emph{universal} $\beta$-function: although this function may depend on the scheme chosen to renormalise the theory, it is unique within a given scheme.% \protect\cite{Celmaster:1979km}.  
%-In MOM, a scheme means choosing a particular vertex with which to define the coupling. A different vertex choice means a different scheme.  The relation between schemes is fixed by the ST identities.  This means that once a scheme is chosen, MOM cannot be used independently to define all renormalisation constants.  Once a subset is chosen, the others are completely determined. 
%Of course, the behaviour of the $\beta$-function on the perturbative domain is well known.}  

This is a well-posed problem whose solution is an elemental goal of modern hadron physics.  Through the gap and Bethe-Salpeter equations (BSEs) the pointwise behaviour of the $\beta$-function determines the pattern of chiral symmetry breaking.  Moreover, the fact that these and other Dyson-Schwinger equations (DSEs) \cite{Roberts:1994dr,Roberts:2007jh,Roberts:2007ji} connect the $\beta$-function to experimental observables entails, e.g., that comparison between computations and observations of the hadron mass spectrum can be used to chart the $\beta$-function's long-range behaviour.  In order to realise this goal a nonperturbative symmetry-preserving DSE truncation is necessary.  Steady quantitative progress can be made with a scheme that is systematically improvable \cite{Munczek:1994zz,Bender:1996bb}. On the other hand, one anticipates that significant qualitative advances could be made with symmetry-preserving kernel \emph{Ans\"atze} that express important additional nonperturbative effects, which are difficult to capture in any finite sum of contributions.  Hitherto no such Ansatz has been available.  We remedy that.

The Poincar\'e covariant bound-state problem is most easily formulated for mesons.  One must first solve the gap equation ($f$ labels the quark flavour):%\footnote{In our Euclidean metric:  $\{\gamma_\mu,\gamma_\nu\} = 2\delta_{\mu\nu}$; $\gamma_\mu^\dagger = \gamma_\mu$; $\gamma_5= \gamma_4\gamma_1\gamma_2\gamma_3$; $a \cdot b = \sum_{i=1}^4 a_i b_i$; and $P_\mu$ timelike $\Rightarrow$ $P^2<0$.}
\begin{eqnarray}
\nonumber \lefteqn{S_f(p)^{-1}= Z_2 \,(i\gamma\cdot p + m_f^{\rm bm})}\\
&&+ Z_1 \int^\Lambda_q\!\! g^2 D_{\mu\nu}(p-q) \frac{\lambda^a}{2}\gamma_\mu S_f(q) \frac{\lambda^a}{2}\Gamma^f_\nu(q,p) ,
\label{gendse}
\end{eqnarray} 
where: $D_{\mu\nu}(k)$ is the dressed-gluon propagator; $\Gamma^f_\nu(q,p)$ is the dressed-quark-gluon vertex; $\int^\Lambda_q$ is a Poincar\'e invariant regularisation of the integral, with $\Lambda$ the regularisation mass-scale; $m^{\rm bm}(\Lambda)$ is the Lagrangian current-quark bare mass; and $Z_{1,2}(\zeta^2,\Lambda^2)$ are respectively the vertex and quark wave function renormalisation constants, with $\zeta$ the renormalisation point.  The gap equation's solution is the dressed-quark propagator, which can be written
\begin{eqnarray} 
%\nonumber 
 S(p)^{-1} & = & i \gamma\cdot p \, A(p^2,\zeta^2) + B(p^2,\zeta^2) \,.%\\ 
% 
%& =& \frac{1}{Z(p^2,\zeta^2)}\left[ i\gamma\cdot p + M(p^2,\zeta^2)\right] . 
\label{sinvp} 
\end{eqnarray}
The mass function $M(p^2)=B(p^2,\zeta^2)/A(p^2,\zeta^2)$ is renormalisation point independent.  
The propagator is obtained from Eq.\,(\ref{gendse}) augmented by a renormalisation condition.  
%Since QCD is asymptotically free, the chiral limit is defined by 
%\begin{equation}
%$Z_2(\zeta^2,\Lambda^2) \, m(\Lambda) \equiv 0\,,\; \forall \Lambda \gg \zeta\,,$
%\end{equation}
%which is equivalent to 
%requiring that the renormalisation-point-invariant current-quark mass is zero; i.e., $\hat m = 0$ \cite{Maris:1997hd,Maris:1997tm}.  
A mass-independent renormalisation scheme can be implemented by fixing all renormalisation constants in the chiral limit \cite{Chang:2008ec}.

Pseudoscalar and axial-vector mesons appear as poles in the inhomogeneous BSE for the axial-vector vertex, $\Gamma_{5\mu}^{fg}$.  An exact form of that equation is \cite{Bender:2002as,Bhagwat:2004hn}
\begin{eqnarray}
\nonumber
\lefteqn{\Gamma_{5\mu}^{fg}(k;P) = Z_2 \gamma_5\gamma_\mu - \int_q g^2D_{\alpha\beta}(k-q)\, }\\
\nonumber
&& \times \frac{\lambda^a}{2}\,\gamma_\alpha S_f(q_+) \Gamma_{5\mu}^{fg}(q;P) S_g(q_-) \frac{\lambda^a}{2}\,\Gamma_\beta^g(q_-,k_-) \\
&+& \int_q g^2D_{\alpha\beta}(k-q)\, \frac{\lambda^a}{2}\,\gamma_\alpha S_f(q_+) \frac{\lambda^a}{2} \Lambda_{5\mu\beta}^{fg}(k,q;P), \label{genbse}
\end{eqnarray}
where $\Lambda_{5\mu\beta}^{fg}$ is a four-point Schwinger function that is completely defined via the quark self-energy \cite{Munczek:1994zz,Bender:1996bb}.  Owing to Poincar\'e covariance, one can use $q_\pm = q\pm P/2$, etc., without loss of generality.  The pseudoscalar vertex, $\Gamma_5^{fg}(k;P)$, satisfies an analogous equation and has the general form
\begin{eqnarray}
\nonumber
\lefteqn{i\Gamma_{5}^{fg}(k;P) = \gamma_5 \left[ i E_5(k;P) + \gamma\cdot P F_5(k;P) \right.}\\
&& \left. + \gamma\cdot k \, G_5(k;P) + \sigma_{\mu\nu} k_\mu P_\nu H_5(k;P) \right].
\label{genG5}
\end{eqnarray}
%The unamputated vertex, $\chi_{5}^{fg}(k;P)$ $=$ $S_f(k_+)$ $\Gamma_{5}^{fg}(k;P)$ $S_g(k_-)$, is expressed through the same four Dirac matrix structures but with different scalar coefficients, which we denote ${\cal E}_5$ -- ${\cal H}_5$.

In any reliable study of light-quark hadrons the solution of Eq.\,(\ref{genbse}) must satisfy the axial-vector Ward-Takahashi; viz., with $\Gamma_5^{fg}$ the pseudoscalar vertex,
\begin{eqnarray}
\nonumber P_\mu \Gamma_{5\mu}^{fg}(k;P)  &= &S_f^{-1}(k_+) i \gamma_5
+  i \gamma_5 S_g^{-1}(k_-) \\
&& - \, i\,[m_f(\zeta)+m_g(\zeta)] \,\Gamma_5^{fg}(k;P)\,,
\label{avwtim}
\end{eqnarray}
which expresses chiral symmetry and the pattern by which it is broken in QCD.  We have established that the condition 
\begin{eqnarray}
\nonumber P_\mu \Lambda_{5\mu\beta}^{fg}(k,q;P)& =& \Gamma_\beta^f(q_+,k_+) \, i\gamma_5+ i\gamma_5 \, \Gamma_\beta^g(q_-,k_-)\\
&& \rule{-1em}{0ex} -  i [m_f(\zeta)+m_g(\zeta)] \Lambda_{5\beta}^{fg}(k,q;P), \label{LavwtiGamma}  
\end{eqnarray}
where $\Lambda_{5\beta}^{fg}$ is the analogue of $\Lambda_{5\mu\beta}^{fg}$ in the pseudoscalar equation, is necessary and sufficient to ensure the Ward-Takahashi identity is satisfied.  

Consider Eq.\,(\ref{LavwtiGamma}).  Rainbow-ladder is the leading-order term in the DSE truncation of Refs.\,\cite{Munczek:1994zz,Bender:1996bb}.  It corresponds to $\Gamma_\nu^f=\gamma_\nu$, in which case Eq.\,(\ref{LavwtiGamma}) is solved by $\Lambda_{5\mu\beta}^{fg}\equiv 0 \equiv \Lambda_{5\beta}^{fg}$.  This is the solution that indeed provides the rainbow-ladder forms of Eq.\,(\ref{genbse}). Such consistency will be apparent in any valid systematic term-by-term improvement of the rainbow-ladder truncation.  

However, Eq.\,(\ref{LavwtiGamma}) is far more than merely a device for checking a truncation's consistency.  For, just as the vector Ward-Takahashi identity has long been used to build \emph{Ans\"atze} for the dressed-quark-photon vertex (e.g., Refs.\,\cite{Ball:1980ay,Curtis:1990zs,Dong:1994jr}), Eq.\,(\ref{LavwtiGamma}) provides a means by which to construct a symmetry preserving kernel of the BSE that is matched to any reasonable \emph{Ansatz} for the dressed-quark-gluon vertex that appears in the gap equation.  With this powerful capacity Eq.\,(\ref{LavwtiGamma}) realises a longstanding goal.

To illustrate, suppose that in Eq.\,(\ref{gendse}) one employs an \emph{Ansatz} for the quark-gluon vertex which satisfies
\begin{equation}
P_\mu i \Gamma_\mu^f(k_+,k_-) = {\cal B}(P^2)\left[ S_f^{-1}(k_+) - S_f^{-1}(k_-)\right]\,, \label{wtiAnsatz}
\end{equation}
with ${\cal B}$ flavour-independent.  
NB.\ While the true quark-gluon vertex does not satisfy this identity, owing to the form of the Slavnov-Taylor identity which it does satisfy, it is plausible that a solution of Eq.\,(\ref{wtiAnsatz}) can provide a reasonable pointwise approximation to the true vertex.

Given Eq.\,(\ref{wtiAnsatz}), then Eq.\,(\ref{LavwtiGamma}) entails ($l=q-k$)
\begin{equation}
i l_\beta \Lambda_{5\beta}^{fg}(k,q;P) =
{\cal B}(l)^2\left[ \Gamma_{5}^{fg}(q;P) - \Gamma_{5}^{fg}(k;P)\right], \label{L5beta}
\end{equation}
with an analogous equation for $P_\mu l_\beta i\Lambda_{5\mu\beta}^{fg}(k,q;P)$.

This identity can be solved to obtain
\begin{eqnarray}
\Lambda_{5\beta}^{fg}(k,q;P) & := & {\cal B}((k-q)^2)\, \gamma_5\,\overline{ \Lambda}_{\beta}^{fg}(k,q;P) \,, \label{AnsatzL5a}
\end{eqnarray}
with, using Eq.\,(\ref{genG5}), 
\begin{eqnarray}
\nonumber
\lefteqn{
\overline{ \Lambda}_{\beta}^{fg}(k,q;P) =  2 \ell_\beta \, [ i \Delta_{E_5}(q,k;P)+ \gamma\cdot P \Delta_{F_5}(q,k;P) ]}\\
\nonumber
&& +  \gamma_\beta \, \Sigma_{G_5}(q,k;P) +
2 \ell_\beta \,  \gamma\cdot\ell\, \Delta_{G_5}(q,k;P)  + [ \gamma_\beta,\gamma\cdot P]\\
&& \times \Sigma_{H_5}(q,k;P) + 2 \ell_\beta  [ \gamma\cdot\ell ,\gamma\cdot P]  \Delta_{H_5}(q,k;P) \,,
\label{AnsatzL5b}
\end{eqnarray}
where $\ell=(q+k)/2$, $\Sigma_{\Phi}(q,k;P) = [\Phi(q;P)+\Phi(k;P)]/2$ and $\Delta_{\Phi}(q,k;P) = [\Phi(q;P)-\Phi(k;P)]/[q^2-k^2]$.  

Now, given any \emph{Ansatz} for the quark-gluon vertex that satisfies Eq.\,(\ref{wtiAnsatz}), then the pseudoscalar analogue of Eq.\,(\ref{genbse}) and Eqs.\,(\ref{gendse}), (\ref{AnsatzL5a}), (\ref{AnsatzL5b}) provide a symmetry-preserving closed system whose solution predicts the properties of pseudoscalar mesons.  
%This system and its predictions can smoothly be connected with those obtained, e.g., in a rainbow-ladder or kindred symmetry-preserving truncation of the DSEs.  Moreover, the 
The system can be used to anticipate, elucidate and understand the impact on hadron properties of the rich nonperturbative structure expected of the fully-dressed quark-gluon vertex in QCD.

To exemplify, herein we consider ground state pseudoscalar and scalar mesons composed of equal-mass $u$- and $d$-quarks.  Scalar meson properties can be determined from the inhomogeneous BSE for the scalar vertex.  It is straightforward to adapt the discussion already presented to derive the scalar-vertex analogues of, e.g., Eqs.\,(\ref{AnsatzL5a}), (\ref{AnsatzL5b}).  (We are aware of the role played by resonant contributions to the kernel in the scalar channel \cite{Holl:2005st} but they are not pertinent to this discussion.)

To proceed we need only specify the gap equation's kernel because the BSEs are completely defined therefrom.  The kernel is typically rendered by writing
\begin{eqnarray}
\nonumber \lefteqn{Z_1 g^2 D_{\rho \sigma}(p-q) \Gamma_\sigma(q,p)} \\
& =& {\cal G}((p-q)^2) \, D_{\rho\sigma}^{\rm free}(p-q) \Gamma_\sigma(q,p)\,, \label{KernelAnsatz}
\end{eqnarray}
wherein $D_{\rho \sigma}^{\rm free}(\ell)$ is the Landau-gauge free-gauge-boson propagator, ${\cal G}(\ell^2)$ is an interaction model and $\Gamma_\sigma(q,p)$ is a vertex \textit{Ansatz}.  Herein we employ the Ball-Chiu model for the dressed-quark-gluon vertex \cite{Ball:1980ay}:
\begin{eqnarray}
\nonumber 
\lefteqn{i\Gamma_\mu(q,k)  =
i\Sigma_A(q^2,k^2)\,\gamma_\mu}\\
&& +
2 \ell_\mu \left[i\gamma\cdot \ell \,
\Delta_A(q^2,k^2) + \Delta_B(q^2,k^2)\right] \!,
\label{bcvtx}
\end{eqnarray}
where $A, B$ appear in Eq.\,(\ref{sinvp});
and a simplified form of the effective interaction in Refs.\,\cite{Maris:1997tm,Maris:1999nt,Maris:2002mt}:
\begin{equation}
\label{IRGs}
\frac{{\cal G}(\ell^2)}{\ell^2} = \frac{4\pi^2}{\omega^6} \, D\, \ell^2\, {\rm e}^{-\ell^2/\omega^2}.
\end{equation}
NB. Equation~(\ref{bcvtx}) does not restrict us to ${\cal B} \equiv 1$ in Eq.\,(\ref{wtiAnsatz}) because a deviation from one can always be absorbed into the dressed-gluon propagator.  

Equation~(\ref{IRGs}) delivers an ultraviolet finite model gap equation.  Hence, the regularisation mass-scale can be removed to infinity and the renormalisation constants set equal to one.  For comparison we also report results obtained in the rainbow-ladder truncation; namely, with 
\begin{equation}
\label{rainbowV}
\Gamma_\sigma(q,p) = \gamma_\sigma\,.
\end{equation}

The active parameters in Eq.\,(\ref{IRGs}) are $D$ and $\omega$ but they are not independent: a change in $D$ is compensated by an alteration of $\omega$ \cite{Maris:2002mt}.  For $\omega\in[0.3,0.5]\,$GeV, using Eq.\,(\ref{rainbowV}), ground-state pseudoscalar and vector-meson observables are roughly constant if %\footnote{The value of $m_g$ is typical of the mass-scale associated with nonperturbative gluon dynamics.}
%\begin{equation}
%\label{gluonmass}
$\omega D  = (0.8 \, {\rm GeV})^3$.% =: m_g^3$.
%\end{equation}
%Herein, we employ $\omega=0.5\,$GeV.  %Therefore $D=m_g^3/\omega=1.0\,$GeV$^2$ corresponds to what might be called the real-world reference value for Eq.\,(\ref{rainbowV}).

\begin{table}[t]
\caption{  
\emph{Upper panel} -- Selected results.  Current-quark masses: \emph{upper rows} -- 6.4\,MeV; \emph{next two rows} -- 6\,MeV; and \emph{lower rows} -- 5\,MeV.  
Notes: (i) $\langle \bar q q \rangle^0$, $f_\pi^0$ are, respectively, the chiral-limit quark condensate and pion decay constant; and (ii) $D \omega = \frac{1}{4}$ is only slightly above the critical interaction strength for DCSB in the rainbow gap equation \protect\cite{Chang:2008ec}, which explains the values in Row~2.
\emph{Lower panel} -- Comparison between exact result for the pion's chiral-limit leptonic decay constant, Eq.\,(\protect\ref{gmorfpi0}), and two oft used estimation formulae: respectively, Eqs.\,(C.4) and (7.57) of Ref.\,\protect\cite{Roberts:1994dr}, with percentage errors in parentheses. Experimentally, $f_\pi = 0.092\,$GeV.
($\omega=0.5\,$GeV throughout;
$A(0)$, dimensionless; $D$, GeV$^2$; and other entries quoted in GeV.)
\label{Table:Para1} 
}
\begin{center}
\begin{tabular*}%{|c|c|c|c|c|c|c|}\hline
{\hsize}
{|l@{\extracolsep{0ptplus1fil}}
|c@{\extracolsep{0ptplus1fil}}
||l@{\extracolsep{0ptplus1fil}}
|l@{\extracolsep{0ptplus1fil}}
|l@{\extracolsep{0ptplus1fil}}
|l@{\extracolsep{0ptplus1fil}}
|l@{\extracolsep{0ptplus1fil}}
|l@{\extracolsep{0ptplus1fil}}|} \hline
\rule{0em}{3ex} 
Vertex & $D$ & $A(0)$ & $M(0)$ & $-(\langle \bar q q \rangle^0)^{1/3}$ & $f_\pi^0$ & $m_\pi$ & $m_\sigma$ \\\hline
Eq.\,(\ref{rainbowV}), {\rm RL} & $\frac{1}{2}$ & 0.97 & 0.049 & 0.13 & 0.029 & 0.16 & 0.27 \\
Eq.\,(\ref{bcvtx}), {\rm BC} &  & 1.1 & 0.28 & 0.26 & 0.11 & 0.14 & 0.56 \\\hline
Eq.\,(\ref{rainbowV}), {\rm RL} & $\frac{2}{3}$ & 1.1&  0.21&  0.21&  0.071&        0.14&         0.44\\
Eq.\,(\ref{bcvtx}), {\rm BC} &  & 
1.3&       0.44&     0.30&    0.13&       0.14&         0.81\\\hline
Eq.\,(\ref{rainbowV}), {\rm RL} & 1 & 1.3 & 0.40 & 0.25 & 0.091 & 0.14 & 0.64 \\
Eq.\,(\ref{bcvtx}), {\rm BC} &  &  1.8 & 0.62  & 0.36  & 0.16 & 0.13  & 1.1\\\hline
\end{tabular*}
\smallskip

\begin{tabular*}%{|c|c|c|c|c|c|c|}\hline
{\hsize}
{|l@{\extracolsep{0ptplus1fil}}
|c@{\extracolsep{0ptplus1fil}}
||l@{\extracolsep{0ptplus1fil}}
|l@{\extracolsep{0ptplus1fil}}
|l@{\extracolsep{0ptplus1fil}}|} \hline
\rule{0em}{3ex} 
Vertex & $D$ & $f_\pi^0$ & $f_{\pi\,{\rm PS}}^0$ & $f_{\pi\,{\rm CR}}^0$  \\
\hline
Eq.\,(\ref{rainbowV}), {\rm RL} &  $\sfrac{2}{3}$ & 0.071 &     0.063 (10\%)&      0.070 (-0.4\%)\\
Eq.\,(\ref{bcvtx}), {\rm BC} &  &         0.13&     0.10 (25\%)&          0.12 (12\%)\\\hline
\end{tabular*}
\end{center}
\end{table}

We obtain meson masses from the inhomogeneous BSEs following the method in Secs.\,3.1, 3.2 of Ref.\,\cite{Bhagwat:2007rj}, with the results presented in Table~\ref{Table:Para1}.  The vacuum quark condensate reported in that table is obtained from the trace of the chiral-limit dressed-quark propagator \cite{Maris:1997hd,Langfeld:2003ye} and the chiral-limit leptonic decay constant is determined from the Gell-Mann--Oakes--Renner relation:
\begin{equation}
\label{gmorfpi0}
(f_\pi^0)^2 =  \frac{-\langle \bar q q \rangle^0_\zeta}{s_\pi^0(\zeta)}\,,\;
s_\pi^0(\zeta) = \left. m_\pi \frac{d m_\pi}{d m(\zeta)}\right|_{\hat m =0}.
\end{equation}
[Remember, the renormalisation point is removed to infinity when using Eq.\,(\ref{IRGs}).]  Both the condensate and decay constant are order parameters for DCSB.  It is evident that dressing the vertex amplifies this phenomenon.  

\begin{figure}[t]
\vspace*{-5ex}

\centerline{\includegraphics[clip,width=0.43\textwidth]{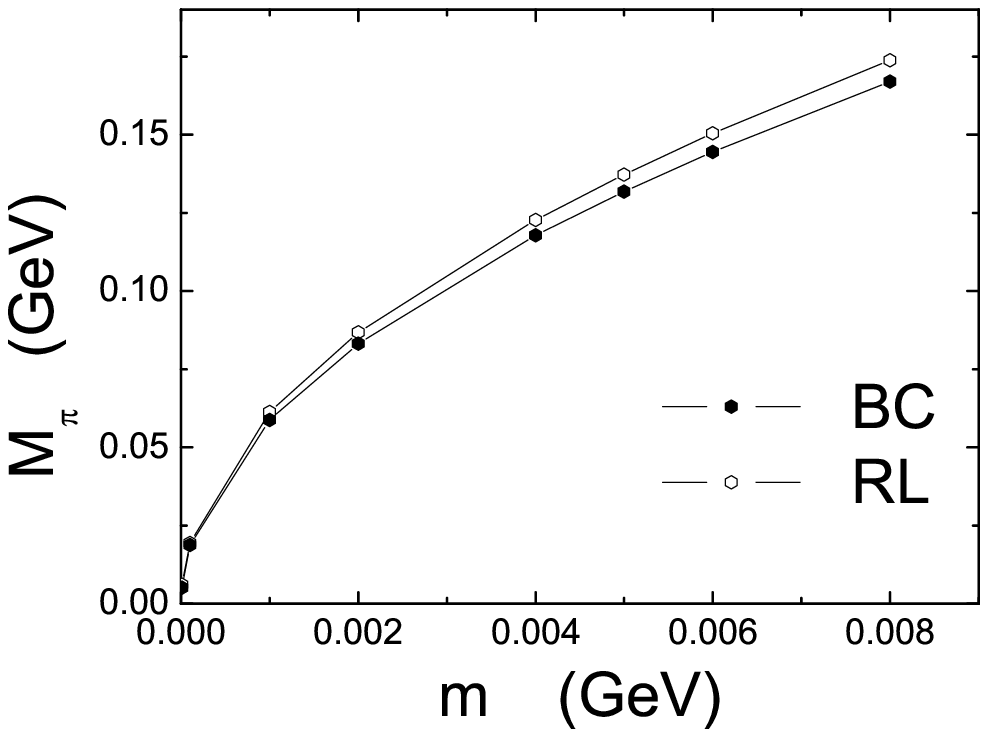}}
\vspace*{-7ex}

\centerline{\includegraphics[clip,width=0.43\textwidth]{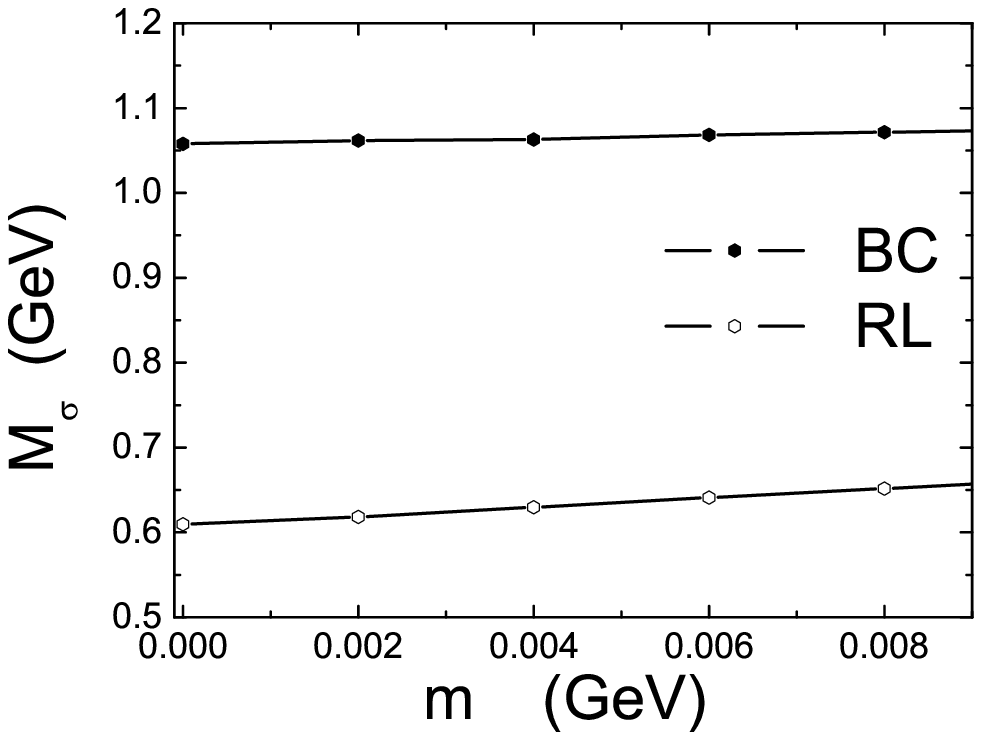}}
\vspace*{-5ex}

\caption{\label{massDlarge} Current-quark-mass dependence of pseudoscalar (upper panel) and scalar (lower) meson masses, obtained with $D=1\,$GeV$^2$.  The Ball-Chiu vertex [BC, Eq.\,(\ref{bcvtx})] result is compared with the rainbow-ladder result [RL, Eq.\,(\ref{rainbowV})].}
\end{figure}

Herein, for the first time, Eq.\,(\ref{gmorfpi0}) can veraciously be used for a truncation whose diagrammatic content is unknown because we have enabled a direct calculation of the current-quark-mass dependence of meson masses obtained with the Ball-Chiu vertex.  That dependence is depicted in Fig.\,\ref{massDlarge} and compared with the rainbow-ladder result.  The $m$-dependence of the pseudoscalar meson's mass provides numerical confirmation that the axial-vector Ward-Takahashi identity is preserved by both the rainbow-ladder truncation and our BC-consistent \emph{Ansatz} for the Bethe-Salpeter kernel.  The figure also shows that the axial-vector Ward-Takahashi identity and DCSB conspire to shield the pion's mass from material variation in response to dressing the quark-gluon vertex \cite{Bhagwat:2004hn,Roberts:2007jh}.

Since our procedure ensures that Eq.\,(\ref{gmorfpi0}) provides a true result for $f_\pi^0$, we can explore the accuracy of two formulae oft used to estimate this quantity.  We find that of Ref.\,\cite{Cahill:1985mh} generally provides the more reliable estimate (see Table~\ref{Table:Para1}).
%--nucl-th/9804062 footnote-1 
The estimation formulae are more reliable in rainbow-ladder truncation because they are derived under the assumption that the bound-state analogues of $F_5$, $G_5$, $H_5$ in Eq.\,(\ref{genG5}) are zero.  
The importance of these amplitudes is signalled by the magnitude of $[A(0)-1]$ \cite{Maris:1997hd}, which, for a given mass-scale in Eq.\,(\ref{IRGs}), is smaller in the rainbow truncation (see Table~\ref{Table:Para1}).  

In the rainbow-ladder DSE truncation, 
%, which is kindred to a mean-field theory, 
%-- Table 1 ... kernel near critical ... M(0) cf. m_sigma is anomalous
using a kernel with realistic interaction strength, one finds \cite{Maris:1997tm,Maris:2000ig,Chang:2006bm,Alkofer:2002bp}
\begin{equation}
\label{epsilonRL}
\varepsilon_\sigma^{\rm RL} := \left.\frac{2 M(0) - m_\sigma }{2 M(0)}\right|_{\rm RL} = (0.3 \pm 0.1)\,.
\end{equation} 
This can be contrasted with the value obtained using our \emph{Ansatz} for the BC-consistent Bethe-Salpeter kernel; viz., 
\begin{equation}
\label{epsilonBC}
\varepsilon_\sigma^{\rm BC} \lesssim 0.1\,.
\end{equation}
Plainly, significant additional repulsion is present in the BC-consistent truncation of the scalar BSE.
%--Can understand nonzero value at increased D ... L.S suppressed by 1/M(0).

Scalar mesons are identified as $^3\!P_0$ states.  This assignment reflects a constituent-quark model perspective, from which a $J^{PC}=0^{++}$ fermion-antifermion bound-state must have the constituents' spins aligned and one unit of constituent orbital angular momentum.  From this viewpoint a scalar is a spin and orbital excitation of a pseudoscalar meson.  Extant studies of realistic corrections to the rainbow-ladder truncation show that they reduce hyperfine splitting \cite{Bhagwat:2004hn}.  Hence, with the comparison between Eqs.\,(\ref{epsilonRL}), (\ref{epsilonBC}) we have a clear indication that in a Poincar\'e covariant treatment the BC-consistent truncation magnifies spin-orbit splitting.  We attribute this to the influence of the quark's dynamically-enhanced scalar self-energy \cite{Roberts:2007ji} in the Bethe-Salpeter kernel.  

We expect this feature to have a material impact on mesons with mass greater than 1\,GeV.  Indeed, \emph{prima facie} it can plausibly overcome a longstanding shortcoming of the rainbow-ladder truncation; viz., that the splitting between vector and axial-vector mesons is too small \cite{Maris:2006ea,Cloet:2007pi}.
This expectation is supported by Refs.\,\cite{Burden:1996nh,Bloch:1999vka} wherein, using a separable Ansatz for the Bethe-Salpeter kernel which depends explicitly on the strength of DCSB, a vector--axial-vector mass-splitting is obtained that is commensurate with experiment.

We presented a Ward-Takahashi identity for the kernel of the pseudovector Bethe-Salpeter equation, Eq.\,(\ref{LavwtiGamma}), and used the identity to construct a symmetry-preserving \emph{Ansatz} for this kernel, Eqs.\,(\ref{AnsatzL5a}), (\ref{AnsatzL5b}), which is consistent with a large class of dressed-quark-gluon vertices whose diagrammatic content cannot be specified.  Although we did not explicitly report formulae, our procedure extends straightforwardly to the vector Bethe-Salpeter equation.  We were therefore able to complete the first exploration of the effect of nonperturbative vertex dressing on the masses of pseudoscalar- and scalar-mesons.  Our results indicate that the dressed-light-quark mass function, which is inextricably connected with dynamical chiral symmetry breaking, acts to magnify spin-orbit splitting in the meson spectrum.  

%\begin{acknowledgments}
%
We thank H.~Chen and A.~Krassnigg for helpful correspondence.
This work was supported by: 
the National Natural Science Foundation of China, contract no.\ 10705002;
the Department of Energy, Office of Nuclear Physics, contract no.\ DE-AC02-06CH11357; 
and the Gordon Godfrey Fund of the School of Physics at the University of New South Wales.
%
%It also benefited from the facilities of ANL's Computing Resource Center.
%\end{acknowledgments}

\bibliography{genvtx}

\begin{thebibliography}{32}
\expandafter\ifx\csname natexlab\endcsname\relax\def\natexlab#1{#1}\fi
\expandafter\ifx\csname bibnamefont\endcsname\relax
  \def\bibnamefont#1{#1}\fi
\expandafter\ifx\csname bibfnamefont\endcsname\relax
  \def\bibfnamefont#1{#1}\fi
\expandafter\ifx\csname citenamefont\endcsname\relax
  \def\citenamefont#1{#1}\fi
\expandafter\ifx\csname url\endcsname\relax
  \def\url#1{\texttt{#1}}\fi
\expandafter\ifx\csname urlprefix\endcsname\relax\def\urlprefix{URL }\fi
\providecommand{\bibinfo}[2]{#2}
\providecommand{\eprint}[2][]{\url{#2}}

\bibitem[{\citenamefont{H{\"o}ll et~al.}(2004)\citenamefont{H{\"o}ll,
  Krassnigg, and Roberts}}]{Holl:2004fr}
\bibinfo{author}{\bibfnamefont{A.}~\bibnamefont{H{\"o}ll}},
  \bibinfo{author}{\bibfnamefont{A.}~\bibnamefont{Krassnigg}},
  \bibnamefont{and} \bibinfo{author}{\bibfnamefont{C.~D.}
  \bibnamefont{Roberts}}, \bibinfo{journal}{Phys. Rev.}
  \textbf{\bibinfo{volume}{C70}}, \bibinfo{pages}{042203(R)}
  (\bibinfo{year}{2004}).

\bibitem[{\citenamefont{H{\"o}ll et~al.}(2005)\citenamefont{H{\"o}ll,
  Krassnigg, Maris, Roberts, and Wright}}]{Holl:2005vu}
\bibinfo{author}{\bibfnamefont{A.}~\bibnamefont{H{\"o}ll}},
  \bibinfo{author}{\bibfnamefont{A.}~\bibnamefont{Krassnigg}},
  \bibinfo{author}{\bibfnamefont{P.}~\bibnamefont{Maris}},
  \bibinfo{author}{\bibfnamefont{C.~D.} \bibnamefont{Roberts}},
  \bibnamefont{and} \bibinfo{author}{\bibfnamefont{S.~V.}
  \bibnamefont{Wright}}, \bibinfo{journal}{Phys. Rev.}
  \textbf{\bibinfo{volume}{C71}}, \bibinfo{pages}{065204}
  (\bibinfo{year}{2005}).

\bibitem[{\citenamefont{Barnes et~al.}(1997)\citenamefont{Barnes, Close, Page,
  and Swanson}}]{Barnes:1996ff}
\bibinfo{author}{\bibfnamefont{T.}~\bibnamefont{Barnes}},
  \bibinfo{author}{\bibfnamefont{F.~E.} \bibnamefont{Close}},
  \bibinfo{author}{\bibfnamefont{P.~R.} \bibnamefont{Page}}, \bibnamefont{and}
  \bibinfo{author}{\bibfnamefont{E.~S.} \bibnamefont{Swanson}},
  \bibinfo{journal}{Phys. Rev.} \textbf{\bibinfo{volume}{D55}},
  \bibinfo{pages}{4157} (\bibinfo{year}{1997}).

\bibitem[{\citenamefont{Burden et~al.}(1997)\citenamefont{Burden, Qian,
  Roberts, Tandy, and Thomson}}]{Burden:1996nh}
\bibinfo{author}{\bibfnamefont{C.~J.} \bibnamefont{Burden}},
  \bibinfo{author}{\bibfnamefont{L.}~\bibnamefont{Qian}},
  \bibinfo{author}{\bibfnamefont{C.~D.} \bibnamefont{Roberts}},
  \bibinfo{author}{\bibfnamefont{P.~C.} \bibnamefont{Tandy}}, \bibnamefont{and}
  \bibinfo{author}{\bibfnamefont{M.~J.} \bibnamefont{Thomson}},
  \bibinfo{journal}{Phys. Rev.} \textbf{\bibinfo{volume}{C55}},
  \bibinfo{pages}{2649} (\bibinfo{year}{1997}).

\bibitem[{\citenamefont{Ackleh et~al.}(1996)\citenamefont{Ackleh, Barnes, and
  Swanson}}]{Ackleh:1996yt}
\bibinfo{author}{\bibfnamefont{E.~S.} \bibnamefont{Ackleh}},
  \bibinfo{author}{\bibfnamefont{T.}~\bibnamefont{Barnes}}, \bibnamefont{and}
  \bibinfo{author}{\bibfnamefont{E.~S.} \bibnamefont{Swanson}},
  \bibinfo{journal}{Phys. Rev.} \textbf{\bibinfo{volume}{D54}},
  \bibinfo{pages}{6811} (\bibinfo{year}{1996}).

\bibitem[{\citenamefont{Bloch et~al.}(1999)\citenamefont{Bloch, Kalinovsky,
  Roberts, and Schmidt}}]{Bloch:1999vka}
\bibinfo{author}{\bibfnamefont{J.~C.~R.} \bibnamefont{Bloch}},
  \bibinfo{author}{\bibfnamefont{Y.~L.} \bibnamefont{Kalinovsky}},
  \bibinfo{author}{\bibfnamefont{C.~D.} \bibnamefont{Roberts}},
  \bibnamefont{and} \bibinfo{author}{\bibfnamefont{S.~M.}
  \bibnamefont{Schmidt}}, \bibinfo{journal}{Phys. Rev.}
  \textbf{\bibinfo{volume}{D60}}, \bibinfo{pages}{111502(R)}
  (\bibinfo{year}{1999}).

\bibitem[{\citenamefont{Roberts et~al.}(2007)\citenamefont{Roberts, Bhagwat,
  H{\"o}ll, and Wright}}]{Roberts:2007jh}
\bibinfo{author}{\bibfnamefont{C.~D.} \bibnamefont{Roberts}},
  \bibinfo{author}{\bibfnamefont{M.~S.} \bibnamefont{Bhagwat}},
  \bibinfo{author}{\bibfnamefont{A.}~\bibnamefont{H{\"o}ll}}, \bibnamefont{and}
  \bibinfo{author}{\bibfnamefont{S.~V.} \bibnamefont{Wright}},
  \bibinfo{journal}{Eur. Phys. J. (Special Topics)}
  \textbf{\bibinfo{volume}{140}}, \bibinfo{pages}{53} (\bibinfo{year}{2007}).

\bibitem[{\citenamefont{Roberts}(2008)}]{Roberts:2007ji}
\bibinfo{author}{\bibfnamefont{C.~D.} \bibnamefont{Roberts}},
  \bibinfo{journal}{Prog. Part. Nucl. Phys.} \textbf{\bibinfo{volume}{61}},
  \bibinfo{pages}{50} (\bibinfo{year}{2008}).

\bibitem[{\citenamefont{Bali et~al.}(2005)\citenamefont{Bali, Neff, Duessel,
  Lippert, and Schilling}}]{Bali:2005fu}
\bibinfo{author}{\bibfnamefont{G.~S.} \bibnamefont{Bali}},
  \bibinfo{author}{\bibfnamefont{H.}~\bibnamefont{Neff}},
  \bibinfo{author}{\bibfnamefont{T.}~\bibnamefont{Duessel}},
  \bibinfo{author}{\bibfnamefont{T.}~\bibnamefont{Lippert}}, \bibnamefont{and}
  \bibinfo{author}{\bibfnamefont{K.}~\bibnamefont{Schilling}}
  (\bibinfo{collaboration}{SESAM}), \bibinfo{journal}{Phys. Rev.}
  \textbf{\bibinfo{volume}{D71}}, \bibinfo{pages}{114513}
  (\bibinfo{year}{2005}).

\bibitem[{\citenamefont{Krein et~al.}(1992)\citenamefont{Krein, Roberts, and
  Williams}}]{Krein:1990sf}
\bibinfo{author}{\bibfnamefont{G.}~\bibnamefont{Krein}},
  \bibinfo{author}{\bibfnamefont{C.~D.} \bibnamefont{Roberts}},
  \bibnamefont{and} \bibinfo{author}{\bibfnamefont{A.~G.}
  \bibnamefont{Williams}}, \bibinfo{journal}{Int. J. Mod. Phys.}
  \textbf{\bibinfo{volume}{A7}}, \bibinfo{pages}{5607} (\bibinfo{year}{1992}).

\bibitem[{\citenamefont{Roberts and Williams}(1994)}]{Roberts:1994dr}
\bibinfo{author}{\bibfnamefont{C.~D.} \bibnamefont{Roberts}} \bibnamefont{and}
  \bibinfo{author}{\bibfnamefont{A.~G.} \bibnamefont{Williams}},
  \bibinfo{journal}{Prog. Part. Nucl. Phys.} \textbf{\bibinfo{volume}{33}},
  \bibinfo{pages}{477} (\bibinfo{year}{1994}).

\bibitem[{\citenamefont{Munczek}(1995)}]{Munczek:1994zz}
\bibinfo{author}{\bibfnamefont{H.~J.} \bibnamefont{Munczek}},
  \bibinfo{journal}{Phys. Rev.} \textbf{\bibinfo{volume}{D52}},
  \bibinfo{pages}{4736} (\bibinfo{year}{1995}).

\bibitem[{\citenamefont{Bender et~al.}(1996)\citenamefont{Bender, Roberts, and
  von Smekal}}]{Bender:1996bb}
\bibinfo{author}{\bibfnamefont{A.}~\bibnamefont{Bender}},
  \bibinfo{author}{\bibfnamefont{C.~D.} \bibnamefont{Roberts}},
  \bibnamefont{and} \bibinfo{author}{\bibfnamefont{L.}~\bibnamefont{von
  Smekal}}, \bibinfo{journal}{Phys. Lett.} \textbf{\bibinfo{volume}{B380}},
  \bibinfo{pages}{7} (\bibinfo{year}{1996}).

\bibitem[{\citenamefont{Chang et~al.}(2009)}]{Chang:2008ec}
\bibinfo{author}{\bibfnamefont{L.}~\bibnamefont{Chang}} \bibnamefont{et~al.},
  \bibinfo{journal}{Phys. Rev.} \textbf{\bibinfo{volume}{C79}},
  \bibinfo{pages}{035209} (\bibinfo{year}{2009}).

\bibitem[{\citenamefont{Bender et~al.}(2002)\citenamefont{Bender, Detmold,
  Roberts, and Thomas}}]{Bender:2002as}
\bibinfo{author}{\bibfnamefont{A.}~\bibnamefont{Bender}},
  \bibinfo{author}{\bibfnamefont{W.}~\bibnamefont{Detmold}},
  \bibinfo{author}{\bibfnamefont{C.~D.} \bibnamefont{Roberts}},
  \bibnamefont{and} \bibinfo{author}{\bibfnamefont{A.~W.}
  \bibnamefont{Thomas}}, \bibinfo{journal}{Phys. Rev.}
  \textbf{\bibinfo{volume}{C65}}, \bibinfo{pages}{065203}
  (\bibinfo{year}{2002}).

\bibitem[{\citenamefont{Bhagwat et~al.}(2004)\citenamefont{Bhagwat, H{\"o}ll,
  Krassnigg, Roberts, and Tandy}}]{Bhagwat:2004hn}
\bibinfo{author}{\bibfnamefont{M.~S.} \bibnamefont{Bhagwat}},
  \bibinfo{author}{\bibfnamefont{A.}~\bibnamefont{H{\"o}ll}},
  \bibinfo{author}{\bibfnamefont{A.}~\bibnamefont{Krassnigg}},
  \bibinfo{author}{\bibfnamefont{C.~D.} \bibnamefont{Roberts}},
  \bibnamefont{and} \bibinfo{author}{\bibfnamefont{P.~C.} \bibnamefont{Tandy}},
  \bibinfo{journal}{Phys. Rev.} \textbf{\bibinfo{volume}{C70}},
  \bibinfo{pages}{035205} (\bibinfo{year}{2004}).

\bibitem[{\citenamefont{Ball and Chiu}(1980)}]{Ball:1980ay}
\bibinfo{author}{\bibfnamefont{J.~S.} \bibnamefont{Ball}} \bibnamefont{and}
  \bibinfo{author}{\bibfnamefont{T.-W.} \bibnamefont{Chiu}},
  \bibinfo{journal}{Phys. Rev.} \textbf{\bibinfo{volume}{D22}},
  \bibinfo{pages}{2542} (\bibinfo{year}{1980}).

\bibitem[{\citenamefont{Curtis and Pennington}(1990)}]{Curtis:1990zs}
\bibinfo{author}{\bibfnamefont{D.~C.} \bibnamefont{Curtis}} \bibnamefont{and}
  \bibinfo{author}{\bibfnamefont{M.~R.} \bibnamefont{Pennington}},
  \bibinfo{journal}{Phys. Rev.} \textbf{\bibinfo{volume}{D42}},
  \bibinfo{pages}{4165} (\bibinfo{year}{1990}).

\bibitem[{\citenamefont{Dong et~al.}(1994)\citenamefont{Dong, Munczek, and
  Roberts}}]{Dong:1994jr}
\bibinfo{author}{\bibfnamefont{Z.-h.} \bibnamefont{Dong}},
  \bibinfo{author}{\bibfnamefont{H.~J.} \bibnamefont{Munczek}},
  \bibnamefont{and} \bibinfo{author}{\bibfnamefont{C.~D.}
  \bibnamefont{Roberts}}, \bibinfo{journal}{Phys. Lett.}
  \textbf{\bibinfo{volume}{B333}}, \bibinfo{pages}{536} (\bibinfo{year}{1994}).

\bibitem[{\citenamefont{H{\"o}ll et~al.}(2006)\citenamefont{H{\"o}ll, Maris,
  Roberts, and Wright}}]{Holl:2005st}
\bibinfo{author}{\bibfnamefont{A.}~\bibnamefont{H{\"o}ll}},
  \bibinfo{author}{\bibfnamefont{P.}~\bibnamefont{Maris}},
  \bibinfo{author}{\bibfnamefont{C.~D.} \bibnamefont{Roberts}},
  \bibnamefont{and} \bibinfo{author}{\bibfnamefont{S.~V.}
  \bibnamefont{Wright}}, \bibinfo{journal}{Nucl. Phys. Proc. Suppl.}
  \textbf{\bibinfo{volume}{161}}, \bibinfo{pages}{87} (\bibinfo{year}{2006}).

\bibitem[{\citenamefont{Maris and Roberts}(1997)}]{Maris:1997tm}
\bibinfo{author}{\bibfnamefont{P.}~\bibnamefont{Maris}} \bibnamefont{and}
  \bibinfo{author}{\bibfnamefont{C.~D.} \bibnamefont{Roberts}},
  \bibinfo{journal}{Phys. Rev.} \textbf{\bibinfo{volume}{C56}},
  \bibinfo{pages}{3369} (\bibinfo{year}{1997}).

\bibitem[{\citenamefont{Maris and Tandy}(1999)}]{Maris:1999nt}
\bibinfo{author}{\bibfnamefont{P.}~\bibnamefont{Maris}} \bibnamefont{and}
  \bibinfo{author}{\bibfnamefont{P.~C.} \bibnamefont{Tandy}},
  \bibinfo{journal}{Phys. Rev.} \textbf{\bibinfo{volume}{C60}},
  \bibinfo{pages}{055214} (\bibinfo{year}{1999}).

\bibitem[{\citenamefont{Maris et~al.}(2003)\citenamefont{Maris, Raya, Roberts,
  and Schmidt}}]{Maris:2002mt}
\bibinfo{author}{\bibfnamefont{P.}~\bibnamefont{Maris}},
  \bibinfo{author}{\bibfnamefont{A.}~\bibnamefont{Raya}},
  \bibinfo{author}{\bibfnamefont{C.~D.} \bibnamefont{Roberts}},
  \bibnamefont{and} \bibinfo{author}{\bibfnamefont{S.~M.}
  \bibnamefont{Schmidt}}, \bibinfo{journal}{Eur. Phys. J.}
  \textbf{\bibinfo{volume}{A18}}, \bibinfo{pages}{231} (\bibinfo{year}{2003}).

\bibitem[{\citenamefont{Bhagwat et~al.}(2007)\citenamefont{Bhagwat, H{\"o}ll,
  Krassnigg, Roberts, and Wright}}]{Bhagwat:2007rj}
\bibinfo{author}{\bibfnamefont{M.~S.} \bibnamefont{Bhagwat}},
  \bibinfo{author}{\bibfnamefont{A.}~\bibnamefont{H{\"o}ll}},
  \bibinfo{author}{\bibfnamefont{A.}~\bibnamefont{Krassnigg}},
  \bibinfo{author}{\bibfnamefont{C.~D.} \bibnamefont{Roberts}},
  \bibnamefont{and} \bibinfo{author}{\bibfnamefont{S.~V.}
  \bibnamefont{Wright}}, \bibinfo{journal}{Few Body Syst.}
  \textbf{\bibinfo{volume}{40}}, \bibinfo{pages}{209} (\bibinfo{year}{2007}).

\bibitem[{\citenamefont{Maris et~al.}(1998)\citenamefont{Maris, Roberts, and
  Tandy}}]{Maris:1997hd}
\bibinfo{author}{\bibfnamefont{P.}~\bibnamefont{Maris}},
  \bibinfo{author}{\bibfnamefont{C.~D.} \bibnamefont{Roberts}},
  \bibnamefont{and} \bibinfo{author}{\bibfnamefont{P.~C.} \bibnamefont{Tandy}},
  \bibinfo{journal}{Phys. Lett.} \textbf{\bibinfo{volume}{B420}},
  \bibinfo{pages}{267} (\bibinfo{year}{1998}).

\bibitem[{\citenamefont{Langfeld et~al.}(2003)\citenamefont{Langfeld, Markum,
  Pullirsch, Roberts, and Schmidt}}]{Langfeld:2003ye}
\bibinfo{author}{\bibfnamefont{K.}~\bibnamefont{Langfeld}},
  \bibinfo{author}{\bibfnamefont{H.}~\bibnamefont{Markum}},
  \bibinfo{author}{\bibfnamefont{R.}~\bibnamefont{Pullirsch}},
  \bibinfo{author}{\bibfnamefont{C.~D.} \bibnamefont{Roberts}},
  \bibnamefont{and} \bibinfo{author}{\bibfnamefont{S.~M.}
  \bibnamefont{Schmidt}}, \bibinfo{journal}{Phys. Rev.}
  \textbf{\bibinfo{volume}{C67}}, \bibinfo{pages}{065206}
  (\bibinfo{year}{2003}).

\bibitem[{\citenamefont{Cahill and Roberts}(1985)}]{Cahill:1985mh}
\bibinfo{author}{\bibfnamefont{R.~T.} \bibnamefont{Cahill}} \bibnamefont{and}
  \bibinfo{author}{\bibfnamefont{C.~D.} \bibnamefont{Roberts}},
  \bibinfo{journal}{Phys. Rev.} \textbf{\bibinfo{volume}{D32}},
  \bibinfo{pages}{2419} (\bibinfo{year}{1985}).

\bibitem[{\citenamefont{Maris et~al.}(2001)\citenamefont{Maris, Roberts,
  Schmidt, and Tandy}}]{Maris:2000ig}
\bibinfo{author}{\bibfnamefont{P.}~\bibnamefont{Maris}},
  \bibinfo{author}{\bibfnamefont{C.~D.} \bibnamefont{Roberts}},
  \bibinfo{author}{\bibfnamefont{S.~M.} \bibnamefont{Schmidt}},
  \bibnamefont{and} \bibinfo{author}{\bibfnamefont{P.~C.} \bibnamefont{Tandy}},
  \bibinfo{journal}{Phys. Rev.} \textbf{\bibinfo{volume}{C63}},
  \bibinfo{pages}{025202} (\bibinfo{year}{2001}).

\bibitem[{\citenamefont{Chang et~al.}(2007)\citenamefont{Chang, Liu, Bhagwat,
  Roberts, and Wright}}]{Chang:2006bm}
\bibinfo{author}{\bibfnamefont{L.}~\bibnamefont{Chang}},
  \bibinfo{author}{\bibfnamefont{Y.-X.} \bibnamefont{Liu}},
  \bibinfo{author}{\bibfnamefont{M.~S.} \bibnamefont{Bhagwat}},
  \bibinfo{author}{\bibfnamefont{C.~D.} \bibnamefont{Roberts}},
  \bibnamefont{and} \bibinfo{author}{\bibfnamefont{S.~V.}
  \bibnamefont{Wright}}, \bibinfo{journal}{Phys. Rev.}
  \textbf{\bibinfo{volume}{C75}}, \bibinfo{pages}{015201}
  (\bibinfo{year}{2007}).

\bibitem[{\citenamefont{Alkofer et~al.}(2002)\citenamefont{Alkofer, Watson, and
  Weigel}}]{Alkofer:2002bp}
\bibinfo{author}{\bibfnamefont{R.}~\bibnamefont{Alkofer}},
  \bibinfo{author}{\bibfnamefont{P.}~\bibnamefont{Watson}}, \bibnamefont{and}
  \bibinfo{author}{\bibfnamefont{H.}~\bibnamefont{Weigel}},
  \bibinfo{journal}{Phys. Rev.} \textbf{\bibinfo{volume}{D65}},
  \bibinfo{pages}{094026} (\bibinfo{year}{2002}).

\bibitem[{\citenamefont{Maris}(2007)}]{Maris:2006ea}
\bibinfo{author}{\bibfnamefont{P.}~\bibnamefont{Maris}}, \bibinfo{journal}{AIP
  Conf. Proc.} \textbf{\bibinfo{volume}{892}}, \bibinfo{pages}{65}
  (\bibinfo{year}{2007}).

\bibitem[{\citenamefont{Cloet et~al.}(2007)\citenamefont{Cloet, Krassnigg, and
  Roberts}}]{Cloet:2007pi}
\bibinfo{author}{\bibfnamefont{I.~C.} \bibnamefont{Cloet}},
  \bibinfo{author}{\bibfnamefont{A.}~\bibnamefont{Krassnigg}},
  \bibnamefont{and} \bibinfo{author}{\bibfnamefont{C.~D.}
  \bibnamefont{Roberts}} (\bibinfo{year}{2007}), \eprint{arXiv:0710.5746
  [nucl-th]}.

\end{thebibliography}

\end{document}